# Genetic algorithms for finding the weight enumerator of binary linear block codes

Saïd NOUH[1*]                     Mostafa BELKASMI[2]

[1*]Phd. Student, Department of Communication Technology, Mohammed V Souissi University, Rabat, National School of Computer Science and Systems Analysis (ENSIAS), 10000, Morocco, E-mail : nouh_ensias@yahoo.fr

[2]Professor, Department of Communication Technology, Mohammed V Souissi University, Rabat, National School of Computer Science and Systems Analysis (ENSIAS), 10000, Morocco, E-mail : belkasmi@ensias.ma

## *ABSTRACT*

In this paper we present a new method for finding the weight enumerator of binary linear block codes by using genetic algorithms. This method consists in finding the binary weight enumerator of the code and its dual and to create from the famous MacWilliams identity a linear system (S) of integer variables for which we add all known information obtained from the structure of the code. The knowledge of some subgroups of the automorphism group, under which the code remains invariant, permits to give powerful restrictions on the solutions of (S) and to approximate the weight enumerator. By applying this method and by using the stability of the Extended Quadratic Residue codes (ERQ) by the Projective Special Linear group $PSL_2$, we find a list of all possible values of the weight enumerators for the two ERQ codes of lengths 192 and 200. We also made a good approximation of the true value for these two enumerators.

**Keywords:** Genetic algorithms, error correcting codes, weights enumerator, Monte Carlo method.

## I. INTRODUCTION

Let C(n,k,d) be a binary linear block code of length n, dimension k and minimum distance d. The error-correcting capability of a code C is directly related to its minimum distance d; if C is linear then d is its lowest non-zero weight; the weight of a word c is the number of non zero elements it contains. The weights enumerator of C is the polynomial $A(x) = \sum_{i=0}^{n} A_i . x^i$, where $A_i$ denotes the number of codewords of weight i in C. The polynomial A provides valuable information about the performance of C and it is one of the keys to obtain an exact expression for the error detection and error correction performance of C [1-2].

C is said to be self dual if it is equal to its dual, and it is said to be formally self dual (*f.s.d*) if it has the same weight enumerator as its dual.

For a *f.s.d* codes, the weight enumerator is given by Gleason's theorem [3]:

$$A(x) = \sum_{i=0}^{\lfloor n/8 \rfloor} K_i (1+x^2)^{n/2-4i} (x^2 - 2x^4 + x^6)^i.$$ Then it is sufficient to find only the $\lfloor n/8 \rfloor + 1$ coefficients $K_i$ to obtain A. The polynomial A is more simplified when the code is doubly even self dual i.e self dual when all weights are divisible by 4. In this case the polynomial A is given by the following expression: $A(x) = \sum_{i=0}^{\lfloor n/24 \rfloor} K_i (1+14x^4+x^8)^{n/8-3i} \{x^4(1-x^4)^4\}^i$ with only $\lfloor n/24 \rfloor + 1$ coefficients.

For random codes, which doesn't have any particularity in their structures, it is known that there isn't algorithms in polynomial time to compute A. The simplest way is to compute it by using an exhaustive research which is feasible only for small codes, i.e codes of dimension of about 45. In this paper we give a new method based on genetic algorithms to have information about the coefficients $A_i$; thus we find a threshold s for which it is sufficient to find only the coefficients $A_i$, with i less than s, to deduce all the others. We call the polynomial

$A'(x,s) = \sum_{i=d}^{s} A_i x^i$ the semi local weight enumerator of degree s. We show that the polynomial

A can be deduced from the polynomial A'. Contrary to the Gleason's theorem, this method is applicable to all binary linear block codes without any restrictions.

For some codes, the wealth of their algebraic structures permits to determine the weight enumerator; so many excellent studies and methods succeed in finding the weight enumerator of quadratic residue codes or their extended codes [4-7] and the weight enumerators are known for all lengths less than or equal to 167. These methods are based on some algorithms for computing the number of codewords of a given weight [4-6].

The minimum weight in quadratic residue code is 27, 27, 31 and 31 respectively for lengths

191, 193, 199 and 223 [8-9]. Here we apply our method in order to find the weight polynomial of two ERQ codes with lengths 192 and 200.

The remainder of this paper is organised as follows. In the next section, we describe briefly

the genetic algorithms. In section 3 we present some algorithms for finding the binary weight

enumerator and some methods for computing the codewords of a given weight. In section 4

we present a method for finding the weight enumerator from the binary weight enumerator and we give the results of its application on some quadratic residue codes. Finally, a conclusion and a possible future direction of this research are outlined in section 5.

## II.   GENETIC ALGORITHMS

Genetic algorithms are heuristic search algorithms premised on the natural selection and

genetic [10-11]. It is defined by:

• Individual or chromosome: a potential solution of the problem, it's a sequence of genes.

• Population: a set of points of the research space.

• Environment: the space of research.

• Fitness function: the function to maximise / minimise.

- Encoding of chromosomes: it depends on the treated problem, the famous known schemes of coding are: binary encoding, permutation encoding, value encoding and tree encoding.

- Operators of evolution:

Selection: it permits to select the best individuals to insert in the intermediate generation.

Crossover: For a pair of parents $(p_1, p_2)$ it permits to create two children $(ch_1; ch_2)$, with a crossover probability $p_c$.

Mutation: The genes of the individual are muted according to the mutation rate $m_r$ and the mutation probability $p_m$.

## III. BINARY WEIGHT ENUMERATOR AND THE COMPUTING OF CODEWORDS OF A GIVEN WEIGHT

1) Binary weight enumerator:

In this section we try to give an answer for the following question:

Let w an integer less than the length of a code C, is there a codeword of weight w in C? Thus if C is linear then we can deduce its minimum distance.

We define the binary enumerator of C by the polynomial: $P(X) = \sum_{i=0}^{n} P_i X^i$ with $P_i = 1$ if C contains a codeword of weight i and $P_i = 0$ if there is no codeword of weight i in C.

### 1.1) Description of the WGA algorithm

In order to use genetic algorithms, in our work, we use binary encoding which consists to treat an individual as a binary sequence. The WGA algorithm permits to verify if a weight is in the code or no and we propose two variants of this algorithm. The first consists in starting with a generation of information vectors of a random weights and lengths equals to the dimension of the code and to try to converge toward a vector I of which the weight of the codeword C(I), obtained by coding I has weight equal to w. The second variant consists in starting with a generation of words with length equals to the length of the code and weights equal to w and to try to converge toward a codeword; all individuals keep the same weight w.

In the sequel of this paper, we use the following notations:

$N_i$ the cardinal of the population; Ng the number of generations; $N_e$ the number of elites (better parents); Ngmax the maximum number of generations and C(I) is the codeword obtained by coding the information vector I.

Algorithm A1: Let w be the objective weight, the WGA algorithm works as follows:

---

*Inputs: w, P;*

1. *Generate the initial population, of $N_i$ individuals; each individual is a word of length k and of a random weight;    Ng $\leftarrow 1$;*

2. *While ($P_w \neq 1$ and Ng < Ngmax):*

2.1 *Compute the fitness of each individual:*

   *fitness (individual) $\leftarrow |weight(C(individual))$-w$|$;    $P_{W(C(individual))} \leftarrow 1$*

2.2 *Sort the population by increasing order of the fitness.*

2.3 *Copy the best $N_e$ individuals (of small fitness) in the intermediate population.*

2.4 *For i = $N_e$ to $N_i$ :*

   3.4.1. *Select a couple of parents ($p_1,p_2$)*

   3.4.2. *Cross $p_1$ and $p_2$ to generate $ch_1$ and $ch_2$ ;  Mute $ch_1$ and $ch_2$*

   3.4.3. *f1 $\leftarrow$ weight(C($ch_1$));  f2 $\leftarrow$ weight(C($ch_2$)); $P_{f1} \leftarrow 1$, $P_{f2} \leftarrow 1$*

   3.4.4. *if (f1 < f2) then insert $ch_1$ in the intermediate population else insert $ch_2$. end if;*

   *End for ;*

3.5 *Ng $\leftarrow$ Ng + 1.*

*End while;*

*Outputs: P*

---

*Algorithm A2:* Let w be the objective weight. The WGA algorithm works as follows:

---

Inputs: w, P

1. *Generate an initial population, of $N_i$ individuals, each individual is a word of length n and weight w and is not necessarily a codeword.*

2. *$Ng \leftarrow 1$*

3. *While ($P_w \neq 1$ and $Ng < Ngmax$):*

3.1 *Compute the fitness of each individual:*

   *- D=D(individual) $\leftarrow$ the nearest codeword to the individual, decided by a hard decision or a soft decision decoder.*

   *- fitness (individual) $\leftarrow$ |weight(D)-w |,*

   *- $P_{W(D)} \leftarrow 1$*

3.2 *Sort the population by increasing order of fitness.*

3.3 *Copy the best $N_e$ individuals (of small fitness) in the intermediate population.*

3.4 *For $i=N_e$ to $N_i$ :*

   3.4.1. *Select an individual $indiv_1$ among the best $N_e$ individuals. Create and mute $indiv_2$ as follow:    - $indiv_2 \leftarrow indiv_1$*

       *- Generate randomly a subset $\Gamma$ : $\Gamma \cap \{1,2,3,...,n\}=\Gamma$*

       *- Generate a random permutation $\sigma$ over $\Gamma$ :  $\sigma(\Gamma)=\Gamma$.*

       *- For j in $\Gamma$ :  $indiv_2$ [j]$\leftarrow indiv_1[\sigma(j)]$ end for;*

   3.4.2. *Among $indiv_1$ and $indiv_2$ select the word with the small fitness and insert it in the intermediate population.*

   *End for.*

3.5 *$Ng \leftarrow Ng + 1$.*

Outputs: P

---

**Remark R1:** For the A1 algorithm, in the case where the coding is systematic, to generate the initial population it is preferable that the weight of each individual is lower than w+1.

1.2)    Description of the BEGA algorithm:

The BEGA algorithm permits to find all coefficients of the binary weight enumerator P, by using the WGA algorithm, The BEGA algorithm works as follows:

---

*1.  For i=0 to n   $P_i \leftarrow 0$ end for;*

*2.  For i=0 to n   if ($P_i = 0$ ) then execute WGA(P,i) end if; end for;*

---

1.3)    Test and validation of the BEGA algorithm:

To validate the BEGA algorithm we tested it on some codes with known weight enumerators.

We use WGA algorithm with following parameters:

- Selection: we use the selection by truncation; we select randomly two individuals, from the best current parents.

- Crossover:  we chose one point from the individual.

- Ngmax=100; $N_i$=1000; $N_e$=500; $p_c$= 0.9; $p_m$=0.15 and $m_r$=0.25.

We choose the two codes:  BCH (255, 207) and RQ (167, 84).

For the BCH code we used the A2 algorithm and the Berlekamp-Massey decoder [12]. For the Quadratic Residue code, we used the A1 algorithm.

 The results coincide with those found from the exact enumerator of the two codes [13], [6].

2)  Computing the number of codewords of a given weight

Generally the weight enumerator A of a linear code has a form comparable to the binomial distribution, and then the number $A_w$ of codewords of weight w grows with w. To find $A_w$ there are three following methods: M1 for linear codes, M2 for half rate codes and M3 for cyclic codes.

M1 method: if the coding is systematic then to compute $A_w$ it is sufficient to encode all information vectors with weight less than or equal to w. This method requires then to enumerate $\sum_{i=1}^{w}\binom{k}{i}$ codewords.

M2 method : suppose that C is a half rate code, i.e. the dimension $k=\dfrac{n}{2}$, this method begins by obtaining two generator matrices $G'=(I',A')$ and $G''=(A'',I')$ for C having disjoint sets of full rank, i.e. the two diagonal matrices I' and I'' each have rank k. if C is self dual then G' and G'' always exist. This method requires the enumeration of $2.\sum_{i=0}^{w/2-1}\binom{n/2}{i}+\binom{n/2}{w/2}$ codewords

[4] and it is simplified in [18] to the enumeration of only $\sum_{i=0}^{w/2}\binom{n/2}{i}$ codewords, for quadratic double circulant self dual codes and for double circulant *f.s.d* codes.

M3 method : supposing that C is a cyclic code. We have the following Chen's theorem [14]:

Let c be a codeword of C of weight w. Then there exists a cyclic shift of c with exactly $r=\left\lfloor \dfrac{k.w}{n}\right\rfloor$ nonzero coordinates among its first k coordinates.

Thus, let L be an initially empty list. To find the number of codewords of weight w, we encode all information vectors v of weight r. For each information vector v if the corresponding codeword is of weight w then we add all its shifts to L and we should guarantee no duplication of any codeword in L.

3) Approximating the number of codewords of a given weight

In [15] Sidel'nikov has proved that the weight enumerator of some binary primitive BCH codes can be given by the approximate following formula: $A_j=2^{-mt}\binom{n}{j}.(1+R_j)$ where $\left|R_j\right| \le K.n^{-0.1}$ and K is a constant. n is the code length in the form $2^m$-1 and t is the error correction capability of the code. In [16] Kasami et al, have given an approximation of the weight enumerator of linear codes with the binomial distribution by using a generalisation of the Sidel'nikov result.

In the case when C has a large length and dimension, the M1, M2 and M3 methods become very complex and unsuitable. If the weight enumerator A is unknown then it is interesting to give, at least, a good approximation of A. In this sub-section we apply a Monte Carlo method [17] in order to get an approximation of A.

In the general case, the Monte Carlo method consists in formulating a game of chance or a stochastic process which produces a random variable whose expected value is the solution of a certain problem. An approximation to the expected value is obtained by means of sampling from the resulting distribution. This method is used for example in the evaluation of some integrals [17].

In order to have a good approximation of $A_w$, we propose the following probabilistic method, which is a variant of Monte Carlo method.

M4 method: Let $C_w$ be the set of all codewords of weight w and $\Gamma$ a subset of $C_w$. We define the dominance rate of $C_w$ related to $\Gamma$ by: $R(\Gamma) = \dfrac{\left|C_w\right|}{\left|\Gamma\right|}$ With symbol $\left|.\right|$ denoting the cardinal.

The WGA algorithm allows finding a codeword of weight w in the code C. By reiterating this algorithm we can find a set S1 of codewords of weight w, and by using the automorphism group of C on the set S1 we find a large set S2 of codewords with the same weight w. S2 may contain multiple copies of some codewords. Let S3 be the largest subset of S2 which doesn't contain any duplicated codeword. Then $A_w$ is approximated by $\left|S3\right|.R(S3)$.

**Remark R2:** In the practice the dominance rate R(S3) is evaluated statistically as following:

1. $s \leftarrow 1$; fix a number $i_{max}$ to a large value.

2. For i from 1 to $i_{max}$ do: - Find a codeword c of weight w by WGA algorithm.

   - If c∈S2 then $s \leftarrow s + 1$ end if.

   end for.

3. Approximate $R(S3)$ by $(1 + i_{max}) / s$.

We find the value of $i_{max}$ statistically; it is the value of i for which the dominance rate R(S3) becomes relatively invariant.

**Remark R3:** In the practice, $|S3|$ is also evaluated statistically as following:

1. $t \leftarrow 0$; For j from 1 to $j_{max}$ do: - Randomly choose a codeword c from S2.

   - Compute the number v of the copies of c in S2;

   - $t \leftarrow t + v$

   end for.

2. Approximate $|S3|$ by $|S2|.j_{max} / t$.

We find the value of $j_{max}$ statistically; it is the value of j for which the rate $j_{max} / t$ becomes relatively invariant.

## IV. FROM THE BINARY WEIGHT ENUMERATOR TO THE EXACT WEIGHT ENUMERATOR

1) Description of the method: Let A and B, respectively the weight enumerators of C and its dual, P and Q respectively the binary weight enumerators of C and its dual, which we find by the BEGA algorithm. We have the following equalities:

MacWilliams-identity [19] : $\forall j \leq n : A_j = 2^{-k} \sum_{i=0}^{n} B_i \sum_{l=0}^{j} (-1)^l \binom{i}{l}\binom{n-i}{j-l}$ (1)

This permits to pass from A to B and vice versa.

$\forall j \leq n : P_j = 0 \Leftrightarrow A_j = 0 \quad and \quad Q_j = 0 \Leftrightarrow B_j = 0$ (2)

From (1) and (2) we construct a linear system (S) of integer variables $A_i$. The resolution of (S) permits to find a threshold s for which it is sufficient to find the semi local weight enumerator $A'(x,s)$ and to deduce A. To obtain $A'(x,s)$ we use M1, M2 or M3 methods. If it is very difficult to apply one of the three methods, it is possible to use the probabilistic method M4.

2)  Application on the weight enumerators of quadratic residue codes:

2.1)    Quadratic residue codes and their extensions

If n is a prime and $n \equiv \pm 1 (mod\, 8)$ ,the Quadratic Residue code $QR(n) = QR\big(n,(n+1)/2,d\big)$ is

a   cyclic   code   with   a   generator   polynomial   $g(x) = \prod_{i \in Q}(x - \beta^i)$ ,   where

$Q = \{ j^2 \bmod n : 1 \leq j \leq n-1 \}$ is the collection of all nonzero quadratic residue integers modulo n and $\beta$ is a primitive $n^{th}$ root of unity in GF($2^m$), where m is the smallest positive integer such that n divides $2^m$ - 1. A $QR(n)$ code, where d is odd, can be extended to a $EQR(n) = EQR\big(n+1,(n+1)/2,d+1\big)$ code whose codewords are obtained by adjoining a parity-check bit to a fixed position $\infty$ of every codeword of the $QR(n)$ code. It is well known that in the binary case all EQR codes, with lengths which are a multiple of 8, are doubly even self-dual; and all EQR codes with lengths not a multiple of 8 are formally self-dual.

Let A and E be, respectively the weight enumerator of the $QR(n)$ and the $EQR(n)$ codes.

By the Pless identity [20] we have: for j $\leq$ (n-1)/2 : $2j.A_{2j} = \big(n-(2j-1)\big).A_{2j-1}$ (3)

By definition of EQR codes and (3) we have: for $j \leq \dfrac{n-1}{2} : E_{2j} = \dfrac{n+1}{n+1-2j} A_{2j} = \dfrac{n+1}{2j} A_{2j-1}$  (4)

## 2.2) The projective special linear group PSL$_2$(n)

For a prime $n \equiv \pm 1 (\mathrm{mod}\, 8)$, the set of permutations over $\{0,1,2,\ldots,\text{n-1},\infty\}$, of the form $y \to (ay+b)/cy+d$ where a,b,c and d are elements of GF(n) verifying : ad-bc=1 form a group called the projective special linear group G=PSL$_2$(n), of order $|G| = n.(n^2-1)/2$.

PSL$_2$(n) can be generated by the three following permutations [19]:

$S : y \to y+1;\ V : y \to \rho^2 y;\ T : y \to -1/y.$  where $\rho$ is a primitive element of GF(n).

For all values of n, the binary EQR(n) code is invariant under G [19].

**Remark R4:** Let be c a codeword of weight w and length n+1 from the $\mathrm{EQR}(n)$ code with $c(\infty)=1$, then the word $c'$ of length n and weight w-1, obtained from c by: $\forall j \neq \infty : c_i^{'} = c_j$ is an element of the QR(n) code.

## 2.3) Congruence of the Number of Codewords of a Given Weight

In [7], Mykkeltveit et al. have demonstrated that it is possible to compute the weight enumerator E of the binary EQR code, modulo $|G|$ as follows:

i.  Factor $|G|$ in prime numbers $|G| = \prod_{i=1}^{l} q_i^{m_i}$, where q$_i$ are prime numbers and m$_i$ is the highest power of q$_i$ that divides $|G|$.

ii. For each divisor $q_i \neq 2$ :

   a) Find a permutation g$_i$ of order q$_i$ from G, g$_i$ is a generator of a group S$_i$ called a Sylow q$_i$-subgroup of G.

   b) Find $E^{q_i}$ the weight enumerator of the subcode C$_i$ fixed by g$_i$.

iii. For the divisor $q_i = 2$ :

a) Find the highest integer m such that $2^m$ divide $\dfrac{(n+1)}{2}$ or $\dfrac{(n+1)}{2}$ .

b) Find two permutations a and b verifying : a∈G and b∈G, $a^{2^m} = 1, b^2 = 1, bab = a^{-1}$.

c) Find $F^2$ the weight enumerator of the subcode $C^2$ fixed by: $H_2 = \{1, a^{2^{m-1}}\}$ .

d) Find $F^0$ the weight enumerator of the subcode $C^0$ fixed by: $G_4^0 = \{1, a^{2^{m-1}}, b, a^{2^{m-1}}b\}$ .

e) Find $F^1$ the weight enumerator of the subcode $C^1$ fixed by: $G_4^1 = \{1, a^{2^{m-1}}, ab, a^{1+2^{m-1}}b\}$ .

f) Find $E^2$ the weight enumerator of the subcode fixed by $S_2$, a Sylow 2-subgroup of G by : $\forall j \leq n : E_j^2 = (2^m+1).F_j^2 - 2^{m-1}.(F_j^0 + F_j^1)$

iv. For each divisor $q_i$ of $|G|$ and for each integer j less than or equal to n, compute $E_j$ modulo $q_i^{m_i}$ according to the following equality: $E_j \bmod q_i^{m_i} = E_j^{q_i} \bmod q_i^{m_i}$ .

v. For each integer j ≤n, compute $E_j$ modulo $|G|$ by using the Chinese remainder theorem.

This method can be applied to any linear code by using its automorphism group [19].

2.4) Weight enumerators of the QR (191) and EQR (191) codes

Let A and E be, respectively the weight enumerator of the $QR(191)$ and the $EQR(191)$ codes.

By using the BEGA algorithm based on A1 algorithm we find the binary weight enumerators P and Q for the QR(191) and its dual respectively:

$P_i = 1 \Leftrightarrow \{28 \leq i \leq 164 \ and \ (i \bmod 4 = 0 \ or \ i \bmod 4 = 3)\} \ or \ i \in \{0, 192\}$ ;

$Q_i = 1 \Leftrightarrow \{28 \leq i \leq 164 \ and \ i \bmod 4 = 0\} \ or \ i \in \{0, 192\}$ . By (3), we have found the solution A of (S) for the QR(191) code. A and E are related by (4) and we give in Table 1 the weight enumerator E of the EQR(191) code which we deduced from the solution A, with $z_1$ and $z_2$ two unknown integers. E is symmetric then we give only its first half.

**Table 1: The form of the weight enumerator E of the EQR(191) code.**

```
  i : Eᵢ
  0 : 1
 28 : 48 z₁
 32 : 6 z₂
 36 : 69065734464 + 11568 z₁ - 192 z₂
 40 : 16681003659936 - 387072 z₁ + 2976 z₂
 44 : 2638181865286080 + 4662144 z₁ - 29760 z₂
 48 : 260118707412159120 - 30019584 z₁ + 215760 z₂
 52 : 16506204128755716672 + 102079872 z₁ - 1208256 z₂
 56 : 688919563458768198624 - 7108608 z₁ + 5437152 z₂
 60 : 19261567021963529559744 - 2055291840 z₁ - 20195136 z₂
 64 : 366292346792783194741815 + 13670572032 z₁ + 63109800 z₂
 68 : 4798230291291549388046400 - 56511000000 z₁ - 168292800 z₂
 72 : 43753732703694320252103840 + 175210813440 z₁ + 387073440 z₂
 76 : 280144274178089715889150656 - 434619319680 z₁ - 774146880 z₂
 80 : 1268289709189717721455882224 + 890278318080 z₁ + 1354757040 z₂
 84 : 4082464373929527973794806080 - 1533608219520 z₁ - 2084721600 z₂
 88 : 9382224038665793129097020640 + 2246629754880 z₁ + 2828613600 z₂
 92 : 15439604564036779974450436032 - 2818036032480 z₁ - 3394336320 z₂
 96 : 18224832149069836877698945680 + 3037942333440 z₁ + 3606482340 z₂
```

Here we have the semi local weight enumerator $A'(x,s) = 48.z_1.x^{28} + 6.z_2.x^{32}$; the value of the threshold s is equal to 32 and A' contain only two unknowns.

Now we apply the method of Mikkelveit et al. in order to know the Ej modulo |G|.

We have n=191 therefore $|G| = 3483840 = 2^6.3.5.19.191$. Subcodes of the QR(191) code that are invariant under $H_2$, $G_4^0$, $G_4^1$, $S_3$, $S_5$, $S_{19}$ and $S_{191}$ are found and the number of codewords of weight 28 and 32 in these subcodes are then computed. The results are summarised in Table 2, where k denotes the dimension of the corresponding subcode. We used the direct computing method for finding the weight enumerators E' of all fixed subcodes of dimension k≠48. For k=48 we used the method M4, thus we found a set of 144 codewords of weight 28 in the corresponding subcode and we verified by the A1 algorithm that there isn't no codeword of weight 28 in this subcode outside this set. As the same we have found that the number of codewords of weight 32 in this subcode is 5274.

**Table 2: the results of applying the Mykkeltveit's method on the EQR(191) code**

|  | $H_2$ | $G_4^0$ | $G_4^1$ | $S_3$ | $S_5$ | $S_{19}$ | $S_{191}$ |
|---|---|---|---|---|---|---|---|
| dimension | 48 | 25 | 24 | 32 | 20 | 6 | 1 |
| $E_{28}'$ | 144 | 6 | 0 | 0 | 0 | 0 | 0 |
| $E_{32}'$ | 5274 | 30 | 42 | 0 | 19 | 0 | 0 |

By the Chinese remainder theorem: $E_{28} \equiv 870960 \bmod 3483840$ and $E_{32} \equiv 239514 \bmod 3483840$

Then $\exists \eta_1 \in IN : E_{28} = \eta_1.3483840 + 870960$ and $z_1 = \eta_1.72580 + 18145$

And $\exists \eta_2 \in IN : E_{32} = \eta_2.3483840 + 239514$ and $z_2 = \eta_2.580640 + 39919$

For this code it is difficult to use one of the first three methods and we will use the fourth method M4 for approximate $A_{27}$ and $A_{31}$, the number of codewords of weight 27 and 31, respectively, in the QR(191) code. We have: $A_{27} = \frac{7}{48} E_{28}$ and $A_{31} = \frac{1}{6} E_{32}$ then we prefer to approximate $A_{27}$ and $A_{31}$ than to approximate $E_{28}$ and $E_{32}$.

Approximation of $A_{27}$ and $A_{31}$: According to the notations used in the method M4, we give in Table 3 the statistic results obtained by applying this method with the remarks R2 and R3 on the RQ(191) code for w=27 and w=31.

**Table 3: Approximation of $A_{27}$ and $A_{31}$ for the QR(191) code**

| w | $|S2|$ | $|S3|$ | R(S3) | R(S3).$|S3|$ |
|---|---|---|---|---|
| 27 | 127015 | 127015 | 1 | 127015 |
| 31 | 7000000 | 5511811 | 3.57 | 19677165 |

From Table 3 an approximate value of $\eta_1$ is 0 and an approximate value of $\eta_2$ is 34.

By $\eta_1 = 0 \ and \ \eta_2 = 34$ we give in the Tables 4 and 5, respectively the corresponding likelihood weight enumerator of the QR(191) code and the EQR(191) code.

**Table 4: A likelihood weight enumerator A of the QR(191) code**

| i : A$_i$ | i : A$_i$ |
|---|---|
| 0 : 1 | 60 : 13242327027308885969100 |
| 27 : 127015 | 63 : 1220974494297511765388885 |
| 28 : 743945 | 64 : 2441948988595023530777770 |
| 31 : 19781679 | 67 : 16993732266235361140080300 |
| 32 : 98908395 | 68 : 30988570603135070314269000 |
| 35 : 12277041273 | 71 : 164076497679489311255544000 |
| 36 : 53200512183 | 72 : 273460829465815518759240000 |
| 39 : 3486010524000 | 75 : 1108904418529771425382575608 |
| 40 : 13246839991200 | 76 : 1692538323019124807162871928 |
| 43 : 604467819340440 | 79 : 5284540455136129960107141600 |
| 44 : 2033209937781480 | 80 : 7398356637190581944149999824 |
| 47 : 65030607700467120 | 83 : 17860781635639559987833361560 |
| 48 : 195091823101401360 | 84 : 22963862102965148555786077200 |
| 51 : 4470424313241968328 | 87 : 43001860177661517697810444960 |
| 52 : 12035757766420683960 | 88 : 508203802099636118246850676800 |
| 55 : 200934904008351929696 | 91 : 739814385354428174792163580200 |
| 56 : 48798476687742611497600 | 92 : 804146071037421929121916935000 |
| 59 : 6019239557867675440500 | 95 : 911241607459815130865413167000 |

**Table 5: A likelihood weight enumerator E of the EQR(191) code**

| i : E$_i$ | |
|---|---|
| 0 : 1 | 60 : 19261566585176561409600 |
| 28 : 870960 | 64 : 366292348289253529616655 |
| 32 : 118690074 | 68 : 479823028693704314543520 |
| 36 : 65477553456 | 72 : 4375373271453048300147840 |
| 40 : 16732850515200 | 76 : 28014427415488962325454553 |
| 44 : 2637677757121920 | 80 : 126828970923267119042571398 |
| 48 : 260122430801868480 | 84 : 408246437386047085436196928 |
| 52 : 16506182079662652288 | 88 : 938222403876251295224955264 |
| 56 : 688919670885778044672 | 92 : 15439604563918501039140805152 |
| | 96 : 18224832149196302617308263340 |

### 2.5) Weight enumerators of the QR (199) and EQR (199) codes

Let A and E be, respectively the weight enumerator of the $QR(199)$ and the $EQR(199)$ codes. By using the BEGA algorithm based on A1 algorithm we find the binary weight enumerators P and Q for the QR(199) and its dual respectively:

$P_i = 1 \Leftrightarrow \{32 \le i \le 168 \text{ and } (i \bmod 4 = 0 \text{ or } i \bmod 4 = 3)\} \text{ or } i \in \{0, 200\}$;

$Q_i = 1 \Leftrightarrow \{32 \leq i \leq 168 \ and \ i \ mod \ 4 = 0 \} \ or \ i \in \{0, 200\}$. By (3), we have found the solution A of (S) for the QR(199) code. A and E are related by (4) and we give in Table 6 the weight enumerator E of the EQR(199) code which we deduced from the solution A, with z is an integer unknown. E is symmetric then we give only its first half.

**Table 6: The form of the weight enumerator E of the EQR(199) code**

| $i$ : $E_i$ |
|---|
| 32 : 25z |
| 36 : 21005534550-450z |
| 40 : 6467522952660+1225z |
| 44 : 1252975498471200+48800z |
| 48 : 152872620852751800-824600z |
| 52 : 12069364505468120400+7427600z |
| 56 : 630615147670747950200-46927800z |
| 60 : 22215915779698502141280+227986400z |
| 64 : 535999851662996527356550-892437300z |
| 68 : 8973312175360724436541800+2896038600z |
| 72 : 105388467829350995361897825-7941316500z |
| 76 : 876310274663366548170765600+18652452000z |
| 80 : 5197894915757311013178267720-37900941000z |
| 84 : 22129281942550350836000132400+67117542000z |
| 88 : 67949637583204730713462120200-104150049000z |
| 92 : 151037779970268049961942408800+142175052000z |
| 96 : 243659108313146247784654076100-171190052250z |
| 100 :285720732951827690430040227204+182092000500z |

We have n=199 therefore $|G| = 3940200 = 2^3.3^2.5^2.11.199$.

Subcodes of the EQR(199) code that are invariant under $H_2, G_4^0$, $G_4^1$, $S_3$, $S_5$, $S_{11}$ and $S_{199}$ are obtained and the number of codewords of weight 32 in these subcodes are then computed. The results are tabulated in Table 7, where k denotes the dimension of the corresponding subcode.

**Table 7: The results of applying the Mykkeltveit's method on the EQR(199) code.**

|  | $H_2$ | $G_4^0$ | $G_4^1$ | $S_3$ | $S_5$ | $S_{11}$ | $S_{199}$ |
|---|---|---|---|---|---|---|---|
| dimension | 50 | 25 | 26 | 34 | 20 | 10 | 1 |
| $E_{32}^{'}$ | 2675 | 33 | 15 | 165 | 0 | 0 | 0 |

We used the direct computing method for finding the weight enumerators E' of all fixed subcodes of dimension k≠50. For k=50 we used the method M4, thus we found a set of 2675 codewords of weight 32 in the corresponding subcode and we verified by the A1 algorithm that there isn't any codeword of weight 32 in this subcode outside this set.

$$E_{32}^{'} \equiv 7 \bmod 8, \ E_{32}^{'} \equiv 3 \bmod 9, \ E_{32}^{'} \equiv 0 \bmod 25, \ E_{32}^{'} \equiv 0 \bmod 11, \ E_{32}^{'} \equiv 0 \bmod 199$$

By the Chinese remainder theorem we obtain: $E_{32} \equiv 2790975 \bmod 3940200$

Then $\exists \eta_3 \in IN : E_{32} = \eta_3.3940200 + 2790975$ and $z = \eta_3.157608 + 111639$

$A_{35} = 3780996219 - 81z \geq 0$ therefore $\eta_3 \leq 295$ and $A_{31} = 4z \leq 186423996$.

In our previous work [21] we had a difficulty in deciding the true value of $\eta_3$ among the 296 possible values, here we will approximate the value of $\eta_3$ by using the fourth method M4 for finding the value of A31, the number of codewords of weight 31 in the QR(199).

By (3) we have: $A_{31} = \dfrac{4}{25} E_{32}$ then we prefer to approximate A31 than to approximate E32.

Approximation of $A_{31}$: According to the notations used in the method M4, we give in Table 8 the statistic results obtained by applying this method with the remarks R2 and R3 on the RQ(199) code for w=31.

**Table 8: Approximation of A₃₁ for the QR(199) code**

| $|S2|$ | $|S3|$ | R(S3) | $|S3|$. R(S3) |
|--------|--------|-------|---------------|
| 7500000 | 4120879 | 1.639 | 6755539 |

From Table 8 an approximate value of $\eta_3$ is 10. By $\eta_3 = 10$, we give in the Tables 9 and 10, respectively the corresponding likelihood weight enumerator of the QR(199) code and the EQR(199) code.

| $i : A_i$ |
| --- |
| 0 : 1 |
| 31 : 22065120 |
| 32 : 115841880 |
| 35 : 3334177539 |
| 36 : 15189031011 |
| 39 : 1294856079132 |
| 40 : 5179424316528 |
| 43 : 275713832445744 |
| 44 : 977530860489456 |
| 47 : 36688337310783312 |
| 48 : 116179734817480488 |
| 51 : 3138045424329256584 |
| 52 : 8931360053860191816 |
| 55 : 176572168865081742536 |
| 56 : 454042719938781623664 |
| 59 : 6664775111200596219984 |
| 60 : 15551141926134724513296 |
| 63 : 171519950956819999396016 |
| 64 : 364479895783242498716534 |
| 67 : 3050926145054268643282932 |
| 68 : 5922386046281815601666868 |
| 71 : 37939848402796009192540017 |
| 72 : 674486193827484607867377808 |
| 75 : 332997904411178304513943728 |
| 76 : 5433123703550803915753818772 |
| 79 : 20791579662192955241435150888 |
| 80 : 311873694932894328621527 2632 |
| 83 : 92942984160266477796045234808 |
| 84 : 128349835268939421945386575 92 |

**Table 9: A likelihood weight enumerator A of the QR(199) code.**

| $i : A_i$ |
| --- |
| 87 : 298978405363572923477123360 88 |
| 88 : 380517970462729175334520641 12 |
| 91 : 694773787866840705845829256 48 |
| 92 : 815640401184368256773206043 152 |
| 95 : 116956371989856919451149654 128 |
| 96 : 126702736322344996072078791 972 |
| 99 : 142860366476416080445279183 602 |

| $i : E_i$ |
| --- |
| 0 : 1 |
| 32 : 115841880 |
| 36 : 15189031011 |
| 40 : 5179424316528 |
| 44 : 977530860489456 |
| 48 : 116179734817480488 |
| 52 : 8931360053860191816 |
| 56 : 454042719938781623664 |
| 60 : 15551141926134724513296 |
| 64 : 364479895783242498716534 |
| 68 : 5922386046281815601666868 |
| 72 : 674486193827484607867377808 |
| 76 : 5433123703550803915753818 72 |
| 80 : 31187369493289432862152726 32 |
| 84 : 128349835268939421945386575 92 |
| 88 : 380517970462729175334520641 12 |
| 92 : 815640401184368256773206043 152 |
| 96 : 126702736322344996072078791 972 |
| 100 : 142860366476416080445279183 602 |

**Table 10: A likelihood weight enumerator E of the EQR(199) code.**

## V. CONCLUSION AND PERSPECTIVES

In this paper we used genetic algorithms for finding all weights in a linear code C, in particular the minimum weight which coincide with the minimum distance. The knowledge of the weights contained in C and its dual are utilised with the MacWilliams identity combined with the automorphism group of C to give an exact or an approximate value of its weight enumerator A(x). In summary our method permits to find a semi local weight enumerator A'(x,s) which is sufficient for finding A(x) and it is less complex to compute than to compute A itself. In the perspectives we have to apply our method on other linear codes like BCH and LDPC codes.